\documentclass[prd,aps,10pt, twocolumn, superscriptaddress, floatfix, nofootinbib, amsmath, amssymb, preprintnumbers]{revtex4-2}

\usepackage{dcolumn}% Align table columns on decimal point
\usepackage{verbatim}

\usepackage{breakurl}

\usepackage{lmodern}

\pdfoutput=1
\usepackage{graphicx}
\usepackage{bm}
\usepackage{amsmath}
\usepackage{amsfonts}
\usepackage{amssymb}
\usepackage{multirow}
\usepackage[colorlinks,linktocpage,linkcolor=cyan,citecolor=cyan]{hyperref}
\usepackage{adjustbox}
\usepackage{array}
\usepackage[capitalise]{cleveref}
\usepackage{acro}
\usepackage[utf8]{inputenc}
\usepackage{CJKutf8}
\usepackage{amssymb}
\usepackage{subfigure}
\usepackage{tikz}
\usepackage{pgfplots}
\pgfplotsset{compat=newest,every axis plot/.append style={line width=1pt}}

\crefname{figure}{Fig.}{Figs.}
\Crefname{figure}{Fig.}{Figs.}
\def\({\left(}
\def\){\right)}
\def\[{\left[}
\def\]{\right]}

\newcommand{\be}{{\begin{eqnarray}}}
\newcommand{\ee}{{\end{eqnarray}}}

\newcommand{\Beq}{\begin{align}}
\newcommand{\Eeq}{\end{align}}

\DeclareAcronym{SW}{
  short = SW ,
  long = Sachs-Wolfe ,
  short-plural =  ,
}
\DeclareAcronym{BH}{
  short = BH ,
  long = black hole ,
  short-plural = s ,
}
\DeclareAcronym{SNR}{
  short = SNR ,
  long = signal-to-noise ratio ,
  short-plural = s ,
}
\DeclareAcronym{IMRPPv2}{
  short = ,
  long = {\normalsize IMRP}{\footnotesize HENOM}{\normalsize P}v2 ,
  short-plural = ,
}
\DeclareAcronym{SFR}{
  short = SFR ,
  long = star formation rate ,
  short-plural =  ,
}
\DeclareAcronym{IMR}{
  short = IMR ,
  long = inspiral-merger-ringdown ,
  short-plural =  ,
}
\DeclareAcronym{ABH}{
	short = ABH ,
	long  = astrophysical black hole,
  short-plural = s ,
}
\DeclareAcronym{GW}{
  short = GW ,
  long = gravitational wave ,
  short-plural = s ,
}
\DeclareAcronym{SIGW}{
  short = SIGW ,
  long = scalar-induced gravitational wave ,
  short-plural = s ,
}
\DeclareAcronym{GWB}{
  short = GWB ,
  long = gravitational-wave background ,
  short-plural = s ,
}
\DeclareAcronym{CBC}{
  short = CBC ,
  long = compact binary coalescence ,
  short-plural = s ,
}
\DeclareAcronym{BBH}{
  short = BBH ,
  long = binary black hole ,
  short-plural = s ,
}
\DeclareAcronym{PBH}{
  short = PBH ,
  long = primordial black hole ,
  short-plural = s ,
}
\DeclareAcronym{LIGO}{
  short =LIGO ,
  long = Laser Interferometer Gravitational-Wave Observatory ,
  short-plural = ,
}
\DeclareAcronym{LVK}{
  short = LVK ,
  long = {Advanced LIGO, Virgo and KAGRA} ,
  short-plural = ,
}
\DeclareAcronym{ET}{
	short = ET ,
	long  = Einstein Telescope,
  short-plural =  ,
}
\DeclareAcronym{CE}{
	short = CE ,
	long  = Cosmic Explorer,
  short-plural =  ,
}
\DeclareAcronym{LISA}{
	short = LISA ,
	long  = Laser Interferometer Space Antenna,
  short-plural =  ,
}
\DeclareAcronym{BBO}{
	short = BBO ,
	long  = big bang observer,
  short-plural =  ,
}
\DeclareAcronym{DECIGO}{
	short = DECIGO ,
	long  = Deci-hertz Interferometer Gravitational wave Observatory,
  short-plural =  ,
}
\DeclareAcronym{SKA}{
	short = SKA ,
	long  = Square Kilometre Array,
  short-plural =  ,
}
\DeclareAcronym{PTA}{
  short = PTA ,
  long = pulsar timing array ,
  short-plural = s ,
}
\DeclareAcronym{FRW}{
  short = FRW ,
  long = Friedman-Robertson-Walker ,
  short-plural =  ,
}
\DeclareAcronym{CMB}{
  short = CMB ,
  long = cosmic microwave background ,
  short-plural =  ,
}
\DeclareAcronym{RD}{
  short = RD,
  long  = radiation-dominated ,
  short-plural =  ,
}
\DeclareAcronym{MD}{
  short = MD,
  long  = matter-dominated ,
  short-plural =  ,
}
\DeclareAcronym{HD}{
  short = HD,
  long  = Hellings-Downs ,
  short-plural =  ,
}
\DeclareAcronym{SMBH}{
  short = SMBH ,
  long  = supper-massive black hole ,
  short-plural = s ,
}
\DeclareAcronym{SGWB}{
  short = SGWB ,
  long  = stochastic gravitational-wave background ,
  short-plural = s ,
}

\DeclareAcronym{PSD}{
  short = PSD ,
  long  = power spectral density ,
  short-plural = s ,
}
\DeclareAcronym{PDF}{
  short = PDF ,
  long  = probability distribution function ,
  short-plural = s ,
}
\DeclareAcronym{ORF}{
  short = ORF ,
  long  = overlap reduction function ,
  short-plural = s ,
}

\DeclareAcronym{NG15}{
  short = NANOGrav ,
  long  = North American Nanohertz Observatory for Gravitational Waves ,
  short-plural =  ,
}
\DeclareAcronym{CPTA}{
  short = CPTA ,
  long  = Chinese PTA ,
  short-plural =  ,
}
\DeclareAcronym{EPTA}{
  short = EPTA ,
  long  = European PTA ,
  short-plural =  ,
}
\DeclareAcronym{PPTA}{
  short = PPTA ,
  long  = Parkes PTA ,
  short-plural =  ,
}

\begin{document}

\title{Unveiling the Graviton Mass Bounds through Analysis of 2023 Pulsar Timing Array Data Releases}

\author{Sai Wang}
\affiliation{Theoretical Physics Division, Institute of High Energy Physics, Chinese Academy of Sciences, Beijing 100049, the People's Republic of China}
\affiliation{University of Chinese Academy of Sciences, Beijing 100049, the People's Republic of China}

\author{Zhi-Chao Zhao}
\email{Corresponding author: zhaozc@cau.edu.cn}
\affiliation{Department of Applied Physics, College of Science, China Agricultural University,
Qinghua East Road, Beijing 100083, the People's Republic of China}

\begin{abstract} 

Strong evidence for the Helling-Downs correlation curves have been reported by multiple pulsar timing array (PTA) collaborations in middle 2023. In this work, we investigate the graviton mass bounds by analyzing the observational data of the overlap ruduction functions from the NANOGrav 15-year data release and CPTA first data release. The results from our data analysis display the state-of-the-art upper limits on the graviton mass at 90\% confidence level, namely, $m_{g}\lesssim8.6\times10^{-24}\mathrm{eV}$ from NANOGrav and $m_{g}\lesssim3.8\times10^{-23}\mathrm{eV}$ from CPTA. We also study the cosmic-variance limit on the graviton mass bounds, i.e., $\sigma_{m_{g}}^{\mathrm{CV}}=4.8\times10^{-24}\mathrm{eV}\times f/(10~\mathrm{year})^{-1}$, with $f$ being a typical frequency band of PTA observations. This is equivalent to the cosmic-variance limit on the speed of gravitational waves, i.e., $\sigma_{v_{g}}^{\mathrm{CV}}=0.07c$, with $c$ being the speed of light. Moreover, we discuss potential implications of these results for scenarios of ultralight tensor dark matter.

\end{abstract}

\maketitle

\acresetall
%%%%%%%%%%%%%%%%%%%%%%%%%%%%%%%%%%%%%%%%%%%%%%%

\section{Introduction}\label{sec:introduction}

In Einstein's theory of general relativity, the \ac{HD} correlation curves \cite{1983ApJ...265L..39H} have been shown to characterize the \acp{ORF} due to the existence of a \ac{SGWB} of nano-Hertz band that is detectable for ongoing and future programs of \ac{PTA}. 
The specific forms of them are solely determined by a null-mass characteristic of gravitational waves. 
Recently, both the \ac{NG15} \cite{NANOGrav:2023gor} and \ac{CPTA} \cite{Xu:2023wog} Collaborations reported strong evidence for a stochastic signal that is spatially correlated among multiple pulsars. 
In particular, the \ac{HD} correlation curves were claimed to deserve statistical significance of $\sim3\sigma-4\sigma$ by \ac{NG15} and of $4.6\sigma$ by \ac{CPTA}. 
Meanwhile, the \ac{EPTA} \cite{Antoniadis:2023ott} and \ac{PPTA} \cite{Reardon:2023gzh} Collaborations further claimed that their observational results are also compatible with the \ac{HD} correlation curves.

However, for the massive gravity, the spatial correlations are expected to be different from the standard \ac{HD} correlation curves. 
Correspondingly, their theoretical expressions depend on the speed of gravitational waves and hence on the graviton mass, but would recover the results of \ac{HD} correlation curves in the massless limit \cite{Liang:2021bct,Bernardo:2022rif,Bernardo:2023pwt,Bernardo:2023bqx}. 
The theory of massive gravity was first proposed by Fierz and Pauli in 1939 \cite{Fierz:1939ix}. 
Since then, it and its extensions have been comprehensively studied in subsequent decades and multiple upper bounds on the graviton mass have been placed by a large quantity of laboratory and astronomy observations (e.g., see reviews in Ref.~\cite{deRham:2016nuf} and references therein). 
Meanwhile, these theories of massive gravity have been used for interpreting cosmological phenomena, such as an accelerating expansion rate of the late universe without the need for dark energy \cite{Kahniashvili:2014wua,Dubovsky:2004ud}, the observed gravitational effects without the need for additional dark matter \cite{Dubovsky:2004ud}, and so on. 
Moreover, constraining the graviton mass with observations from \ac{PTA} could be helpful to understand the nature of gravity, e.g., its intrinsic symmetry, quantization, etc \cite{Janssen:2014dka}.

During the era of gravitational-wave astronomy, the speed of gravitational waves $v_g$ has been shown by the \ac{LVK} Collaborations to be compatible with the light speed with a precision $|v_g-1|\lesssim10^{-15}$, indicating upper limits on the graviton mass, namely, $m_g\lesssim10^{-23}\mathrm{eV}$ \cite{LIGOScientific:2019fpa,LIGOScientific:2021sio}. 
Similar bounds on the graviton mass have also been claimed by other research groups \cite{Wu:2023pbt,Bernardo:2023mxc}, who focused on analysis of the \ac{NG15} 12.5-year \ac{PTA} data \cite{NANOGrav:2020bcs}. 
Moreover, pulsar timing of binary pulsars PSR B1913+16 and PSR B1534+12 has set an upper limit of $ m_g < 7.6 \times 10^{-20} \mathrm{eV}$ at 90\% confidence level \cite{Finn:2001qi}. 
Forecasting works related to planned space-borne gravitational-wave detectors can be found in Ref.~\cite{Gao:2022hho}.

In this work, we will study the state-of-the-art upper limits on the graviton mass through analysis of the 2023 data releases of \ac{NG15} and \ac{CPTA} programs \cite{NANOGrav:2023gor,Xu:2023wog}.
Though demonstrating strong evidence for the \ac{HD} correlation curves, these observations have not vetoed other alternative correlations. 
As shown in Ref.~\cite{Bernardo:2022rif}, the graviton mass would flatten the spatial correlation curves, since a more-massive graviton has a lower speed. 
An earlier forecasting work can be found in Ref.~\cite{Lee:2010cg}. 
Therefore, we will take into account the specific \acp{ORF} for massive gravitons during our practical analysis of the recent \ac{PTA} datasets from \ac{NG15} \cite{NANOGrav:2023gor} and \ac{CPTA} \cite{Xu:2023wog}. 
The remaining context of the paper is as follows. 
In \cref{sec:theory}, we show the theoretical expressions of \acp{ORF} for massive gravitons, with the \ac{HD} correlation curves being their massless limit. 
In \cref{sec:method}, we show a likelihood method for our data analysis and the resulting bounds on the graviton mass. 
In \cref{sec:conclusion}, we display the concluding remarks and discussion.

\section{Spatial correlation curves for massive gravitons}\label{sec:theory}

For simplicity, we consider the helicity-2 modes without loss of generality \cite{deRham:2014zqa}. 
In principle, we can study the massive gravity via analyzing its Lagrangian. 
However, when we focus solely on the propagation of gravitaitonal waves, an equivalent but simple approach seems to concern the dispersion relation of gravitons, i.e., 
\begin{equation}\label{eq:ekm}
    \omega^{2}=k^{2}+m_{g}^{2}\ , 
\end{equation}
where $m_g$ denotes the graviton mass, and $\omega$ and $k$ stand for the angular frequency and wavenumber, respectively.
The above dispersion relation leads to a definition of the speed of gravitational waves, namely $v_{g}=d\omega/dk$ \cite{LIGOScientific:2021sio}. 
For subluminal gravitons, we can estabilish a relation between the graviton mass and the speed of gravitational waves, namely, 
\begin{equation}\label{eq:vgmg}
    m_{g}\simeq1.31\times10^{-22}\mathrm{eV}\times \frac{f}{\mathrm{year}^{-1}} \times \sqrt{1-v_{g}^{2}} \ ,
\end{equation} 
where $f=\omega/(2\pi)$ is the frequency of gravitational waves, and the light speed is defined as unity, i.e., $c=1$. 
For a typical frequency band of gravitational waves, a lower bound on the speed of gravitational waves can be recast into an upper bound on the graviton mass.

Upon the influence of a \ac{SGWB}, the time of arrivals of radio pulses from two pulsars would be spatially correlated  \cite{Ng:2021waj,Mingarelli:2014xfa,Mingarelli:2013dsa}, with the angular correlation being defined as 
\begin{equation}
    \gamma_{ab}(v_{g},\zeta_{ab}) = \sum_{\ell} \frac{2\ell+1}{4\pi} C_{\ell} P_{\ell}(\cos\zeta_{ab})\ ,
\end{equation}
where the subscript $_{ab}$ stands for the cross correlation between these two pulsars, as labeled by $a$ and $b$, with their angular separation $\zeta_{ab}$, and the angular power spectrum $C_{\ell}$ is defined as
\begin{equation}
    C_{\ell} = \frac{1}{\sqrt{\pi}} J_{\ell}(v_{g},fD_{a}) J_{\ell}^{\ast}(v_{g},fD_{b})\ ,
\end{equation}
where $D_{c}$ denotes a distance to the $c$-th pulsar. 
To simplify the definition, we have introduced a function of the form  
\begin{equation}
    J_{\ell}(v_{g},y) = \sqrt{2}\pi i^{\ell} \sqrt{\frac{(\ell+2)!}{(\ell-2)!}} \int_{0}^{2\pi y v_{g}} \frac{dx}{v_{g}} e^{i x/v_{g}} \frac{j_{\ell}(x)}{x^{2}}\ , 
\end{equation}
where $j_{\ell}(x)$ denotes the spherical Bessel function with $\ell$-th multipole. 
Following Refs.~\cite{Bernardo:2022rif,Bernardo:2022xzl}, we recast the angular correlation $\gamma_{ab}(v_{g},\zeta_{ab})$ into the \acp{ORF} of massive gravitons, i.e., $\Gamma_{ab}(v_{g},\zeta_{ab})$, by normalizing the former such that $\Gamma_{ab}(v_{g}=1,\zeta_{ab}=0^+)=0.5$, where the subscript $_{ab}$ stands for the correlation of $a$-th and $b$-th pulsars. 
It is worthy noting that the \ac{HD} correlation curves would be recovered by the above correlations in the massless limit.

\begin{figure}
    \centering
    \includegraphics[width = \columnwidth]{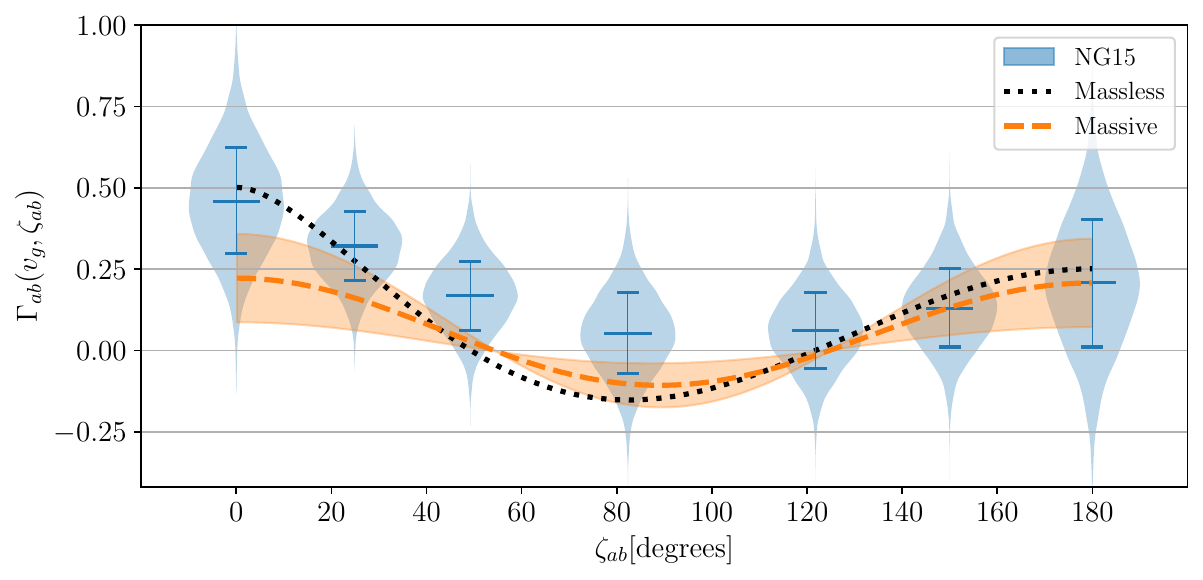}
    \caption{Comparison between the ORFs of massive gravitons (orange dashed curve denotes $m_{g}=7.7\times10^{-24}\mathrm{eV}^{-1}$) and those of massless gravitons. The orange shaded region stands for $1\sigma$ uncertainties due to the cosmic variance. The \ac{NG15} 15-year data points are also shown as shaded violins for comparison. }
    \label{fig:1}
\end{figure}

\begin{figure}
    \centering
    \includegraphics[width = \columnwidth]{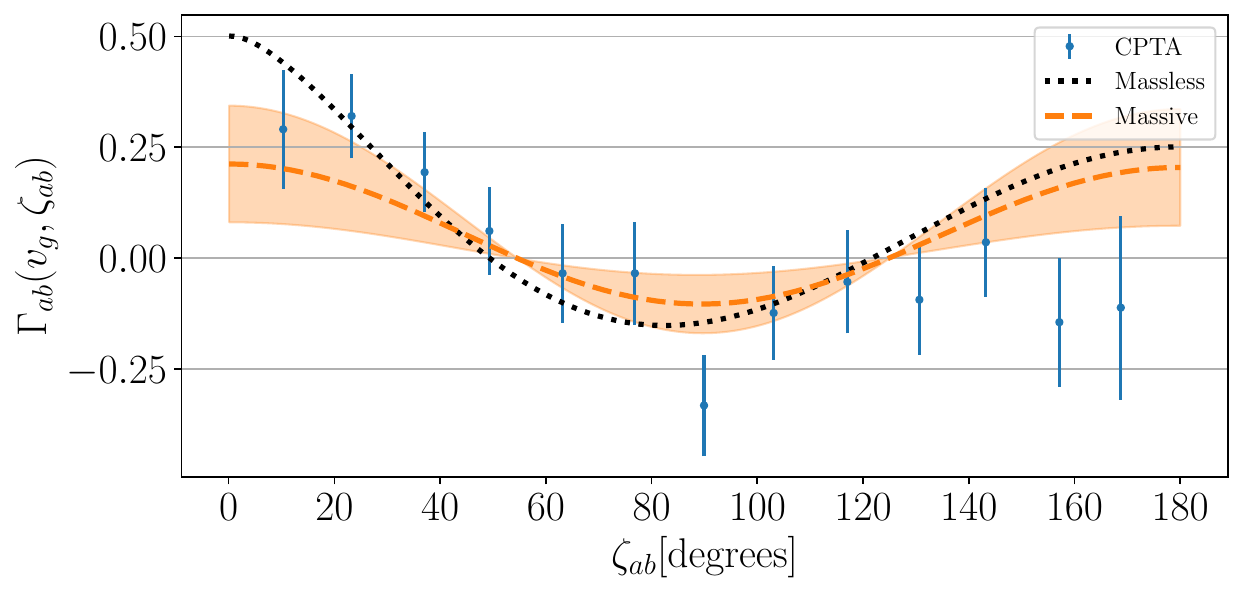}
    \caption{Comparison between the ORFs of massive gravitons (orange dashed curve denotes $m_{g}=3.6\times10^{-23}\mathrm{eV}^{-1}$) and those of massless gravitons. The orange shaded region stands for $1\sigma$ uncertainties due to the cosmic variance. The \ac{CPTA} first data release is also shown as data points with error bars for comparison. }
    \label{fig:2}
\end{figure}

We show the difference between the \acp{ORF} of massive gravitons (orange dashed curves) and those of massless gravitons (black dotted curves) in Figs.~\ref{fig:1} and \ref{fig:2}. 
We depict these \acp{ORF} of massive gravitons by using the best-fit results in the next section. 
The orange shaded regions stand for the $1\sigma$ uncertainties due to the cosmic variance, i.e., $\Delta{C_{\ell}}/{C_{\ell}}=[2/(2\ell+1)]^{1/2}$. 
For comparison, we reproduce the observed \acp{ORF} by \ac{NG15} \cite{NANOGrav:2023gor} (blue shaded violins in Fig.~\ref{fig:1}) and \ac{CPTA} \cite{Xu:2023wog} (blue data points with error bars in Fig.~\ref{fig:2}). 
It seems that the massive gravitons fit better the observed spatial correlation curves than the massless gravitons. 
Our data analysis in the subsequent section would be proved to support such a suspicion\footnote{Besides our present paper, Ref.~\cite{Bernardo:2023zna} supports the same results.}.

\section{Results from data analysis}\label{sec:method}

In the \ac{NG15} 15-year data release, there are 67 individual pulsars with timing baselines longer than three years monitored, implying a total of 2,211 distinct pairs, with each pair having a deterministic angular separation. Based on the previous 12.5-year data release, the authors of Ref.~\cite{NANOGrav:2023gor} have constructed a minimally modeled Bayesian reconstruction of the inter-pulsar correlation pattern via employing a spline interpolation over seven spline-knot positions, i.e., Fig.~1 (d) of the paper. 
In the \ac{CPTA} first data release \cite{Xu:2023wog}, there are 57 individual millisecond pulsars monitored. The $4.6\sigma$ statistical significance of the \ac{HD} correlations between those pulsars has been reported at around $14$ nHz, i.e., Fig.~4 of the data-release paper.

Analyzing the above \ac{NG15} and \ac{CPTA} observational data of the spatial correlations, we can infer the speed of gravitational waves, or equivalently, the graviton mass, by using the log-likelihood as follows 
\begin{equation}\label{likelihood}
-2\ln \mathcal{L}(v_{g}|{\rm D}) = \sum_{\zeta_{ab}} \left( \frac{\Gamma_{ab}(v_{g},\zeta_{ab})-\Gamma^{\rm D}_{ab}}{\sigma^{\rm D}_{ab}} \right)^{2} \ ,
\end{equation}
where $\Gamma^{\rm D}_{ab}$ stands for the observed \acp{ORF} for the binned angular separation $\zeta_{ab}$, $\sigma^{\rm D}_{ab}$ stands for the corresponding $1\sigma$ uncertainty, $\rm D$ denotes the \ac{NG15} or \ac{CPTA} datasets, and the summation runs over all the binned angular separations. 
It is worthy noting that the above uncertainties $\sigma^{\rm D}_{ab}$ also contain contributions from the cosmic variance \cite{Bernardo:2022rif,Bernardo:2022xzl,Allen:2022dzg}, which contributes to an overall error in the measured inter-pulsar correlation or the harmonic space coefficients, $C_l$'s. 
That is $(\sigma^{\rm D}_{ab})^2=(\sigma_m)^2 + (\sigma_{\rm CV})^2$, where $\sigma_m$ denotes the measured error and $\sigma_{\rm CV}$ arises from the cosmic variance that will also be mentioned in the next section. 
In other words, we consider the cosmic variance of the correlation in our data analysis, i.e., the orange regions in Figs. \ref{fig:1} and \ref{fig:2}.

For our data analysis, the parameter to be inferred is the speed of gravitational waves $v_{g}$. 
We let the corresponding prior probability distribution to be uniform, i.e., $v_{g}\in [10^{-2},1]$, and perform Markov-Chain Monte-Carlo samplings by using the publicly available \texttt{cobaya} software \cite{Torrado:2020dgo}. 
We further adopt the publicly available \texttt{PTAfast} package \cite{Bernardo:2022xzl} to compute the \acp{ORF} of massive gravitons. 
The resulting posterior probability distribution functions of $v_{g}$ will be recast into those of $m_{g}$, following the relation in Eq.~(\ref{eq:vgmg}). 
Following the above approach, we will obtain some upper limits on $m_{g}$ at 90\% confidence level and so on.

\begin{figure}
    \centering
    \includegraphics[width = \columnwidth]{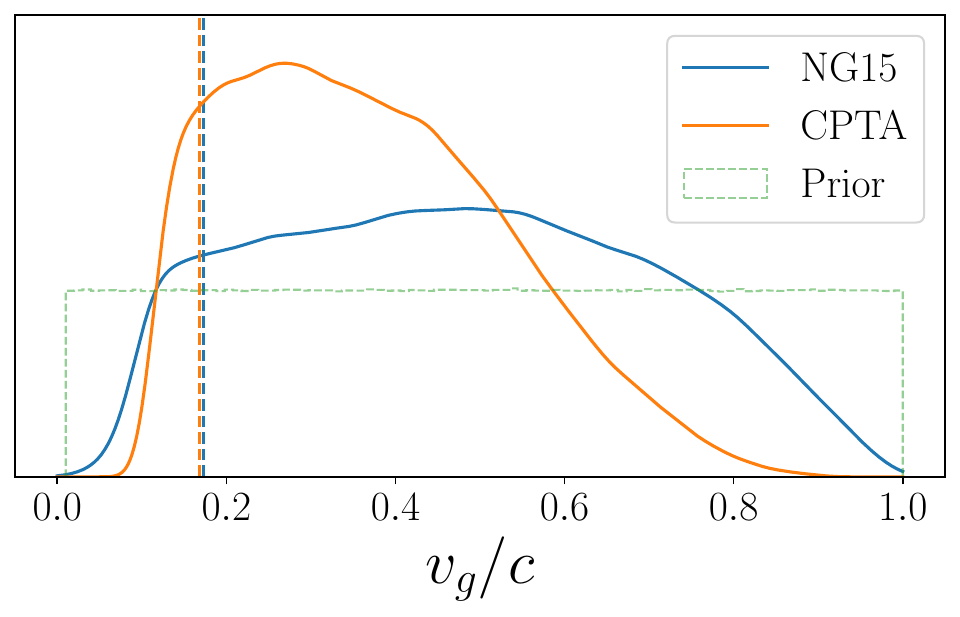}
    \caption{Posteriors of the speed of gravitational waves $v_g$ inferred from the \ac{NG15} 15-year data release and the \ac{CPTA} first data release. The vertical dashed lines stand for the 90\% confident lower limits. }
    \label{fig:3}
\end{figure}

\begin{figure}
    \centering
    \includegraphics[width = \columnwidth]{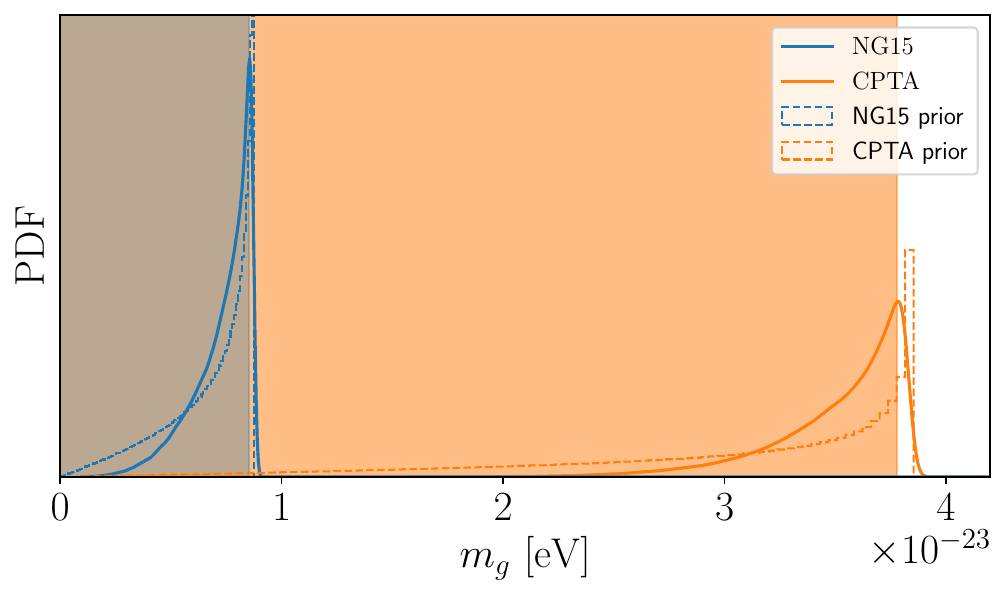}
    \caption{Posteriors of the graviton mass $m_g$ inferred from the \ac{NG15} 15-year data release and the \ac{CPTA} first data release. The shaded regions stand for the allowed parameter space at 90\% confidence level. }
    \label{fig:4}
\end{figure}

Our results are shown as follows. 
Firstly, the posterior probability distributions of $v_g$ are depicted in Fig.~\ref{fig:3}. 
We show the results for \ac{NG15} in a blue curve while those for \ac{CPTA} in an orange curve. 
Correspondingly, we label the 90\% confident lower limits on $v_g$ with vertical dashed lines with the same coloring. 
They are shown to be $v_g\gtrsim0.173$ for \ac{NG15} and $v_g\gtrsim0.168$ for \ac{CPTA}. 
Based on Fig.~\ref{fig:3}, it seems that \ac{CPTA} favors a relatively smaller speed of gravitons, compared with \ac{NG15}. 
Secondly, we further display the posterior probability distributions of $m_g$ in Fig.~\ref{fig:4}. 
The shaded regions stand for the allowed parameter space at 90\% confidence level. 
To be specific, we display the results for \ac{NG15} in the blue shaded region and those for \ac{CPTA} in the orange shaded region. 
Following Eq.~(\ref{eq:vgmg}), in which $f$ is a frequency corresponding to the time span of \ac{PTA} observation, we obtain the upper limits on $m_g$ at 90\% confidence level, namely, 
\begin{equation}
    m_g\lesssim8.6\times10^{-24}\mathrm{eV}
\end{equation} 
from \ac{NG15} and 
\begin{equation}
    m_g\lesssim3.8\times10^{-23}\mathrm{eV}
\end{equation}
from \ac{CPTA}. 
We find that the two bounds are comparable with each other. 
They stand for the state-of-the-art upper limits on the graviton mass. 
Moreover, we notice overlapping posteriors for the \ac{GW} speed, while seemingly distant posteriors for the graviton mass. 
This difference stems from different duty circles or frequency inputs for the \ac{NG15} and \ac{CPTA} data sets. 
In fact, when evaluating Eq.~(\ref{eq:vgmg}), we have used $f=(15 \rm years)^{-1}$ for the former, while $f=(3.4 \rm years)^{-1}$ for the latter.

\begin{figure}
    \centering
    % 第一行的三个子图
    \begin{subfigure}
        \centering
        \includegraphics[width=0.23\textwidth]{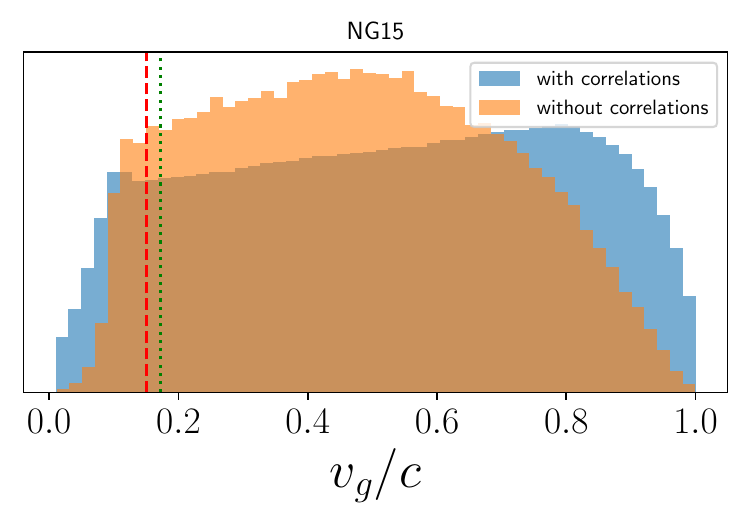}
    \end{subfigure}
    \begin{subfigure}
        \centering
        \includegraphics[width=0.23\textwidth]{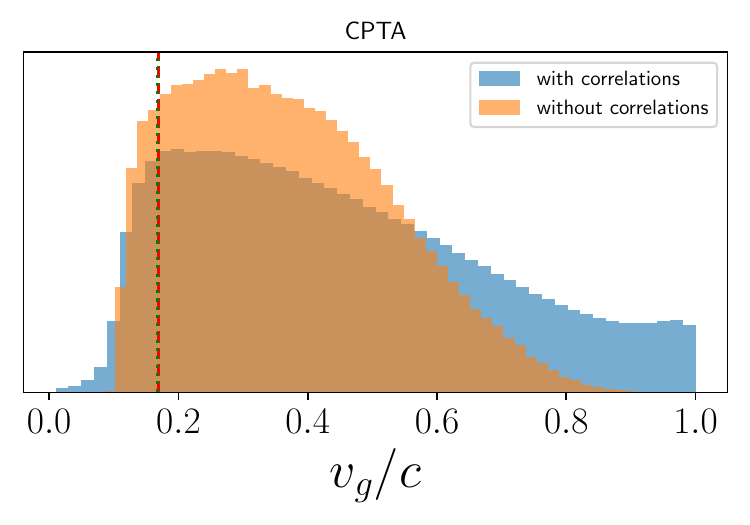}
    \end{subfigure}
    % 如果有第三个子图，应在这里继续添加

    \caption{Comparison between the posteriors with and without monopolar and dipolar correlations inferred from the NG15 (left) and CPTA (right) data sets. The red dashed and green dotted lines represent the 90\% C.L. upper limits for the respective cases.}
    \label{fig:comparison}
\end{figure}

Besides of \ac{SGWB} with characteristic quadrupolar angular patterns, there are two other common signals with different angular correlations. 
The first is associated with errors in our clock standards and has a monopolar correlation. 
The second is due to errors in Solar system ephemerids and has a dipolar correlation. 
To produce robust upper limits on graviton mass, we also compare the data-derived correlation patterns with a combination of these three correlations. 
In other words, the latter can be given by 
\begin{equation}
    \Gamma_{ab}^{\rm eff}(v_{g},\zeta_{ab}) = a + b \times \cos{\zeta_{ab}} + \Gamma_{ab}(v_{g},\zeta_{ab})\, ,
\end{equation}
where $a$ and $b$ are undetermined parameters. 
Fig.~\ref{fig:comparison} demonstrate comparisons between the posteriors of $v_{g}$ with and without monopolar and dipolar correlations for \ac{NG15} and CPTA. 
However, we find few difference between these results of this joint analysis and those of our present analysis.

In Fig.~\ref{fig:4}, we find that the posteriors on the graviton mass are not flat. 
Therefore, it may be more adequate to provide the ``measured'' values of the model parameters. 
At 68\% C.L., they are given by 
\begin{equation}
    m_{g} = 7.71^{+0.81}_{-1.71}\times 10^{-24} \mathrm{eV}
\end{equation}
for \ac{NG15} and 
\begin{equation}
    m_{g} = 3.61^{+0.17}_{-0.37}\times 10^{-23} \mathrm{eV}
\end{equation}
for \ac{CPTA}, respectively. 
The inclination of our results towards the massive graviton hypothesis could arise from several factors that may bias our results. 
For example, the reported signals may not be solely contributed by a \ac{SGWB}, but also by unknown physical processes, e.g., the aforementioned scalar or vector correlations. 
On the other hand, there may also be unknown systematics within the data sets, which should be checked by future observations. 
However, we can not exclude the possibility that the speed of gravitational waves is indeed subluminal. 
It can also be vetoed or confirmed by future observations.

\section{Cosmic-variance limit on the graviton mass bounds}

Our study can be further improved by upcoming \ac{PTA} observations. 
At present, the detection of pulsars is on the rise, with an increasing number of newly-constructed telescopes offering heightened sensitivity and producing data of continually improving quality. 
Concurrently, advancements in pulsar modeling techniques and data processing tools are allowing for more effective noise suppression. 
Moreover, the accumulation of data over extended periods is facilitating an increase in signal-to-noise ratio and enabling observations across diverse frequency bands. 
This has resulted in the acquisition of increasingly powerful tools for \ac{PTA} observations.

However, the cosmic variance, i.e., $\Delta{C_{\ell}}/{C_{\ell}}=[2/(2\ell+1)]^{1/2}$, would inevitably introduce an uncertainty to inferences of the graviton mass \footnote{It is challenging to discuss details of the cosmic variance in this short paper. Here, we provide only a brief review of the concept of cosmic variance. Traditionally, it is popular to consider the cosmic variance in cosmological data analysis, e.g., temperature and polarization anisotropies in the \ac{CMB} \cite{Planck:2018vyg}, etc. Recently, a similar proposal was originally introduced to the community of \ac{PTA} in Refs. \cite{Allen:2022dzg,Allen:2022ksj}. Subsequently, the cosmic variance was further considered in \ac{PTA}-relevant studies of additional polarization modes of gravitational waves in Refs. \cite{Bernardo:2022xzl}. We strongly recommend readers to refer to these papers if they wish to find more details of the cosmic variance. }. 
For example, it can bring about uncertainties on $\Gamma_{ab}({v_{g},\zeta_{ab}})$, as shown as the orange shaded regions in Figs.~\ref{fig:1} and \ref{fig:2}, which could be recast into uncertainties on $m_{g}$, as studied in this section. 
Due to the cosmic variance, we inquire the maximal capability of \ac{PTA} in constraining the graviton mass and whether it is sufficient to challenge theories of gravity beyond general relativity. 
We study this issue in the following.

To investigate the cosmic-variance limit, we consider an ideal \ac{PTA} observation, which does not have instrumental uncertainties. 
In other words, we assume that all the uncertainties arise from the cosmic variance. 
As a first step, we should get the cosmic-variance limit on the speed of gravitational waves. 
For simplicity, we follow an approach of Fisher matrix, i.e., 
\begin{equation}
    F=\int d\zeta_{ab} \left( \frac{1}{\sigma_{ab}} \right)^{2} \left( 
\frac{\partial \Gamma_{ab}(v_{g},\zeta_{ab})}{\partial v_{g}} \right)^{2}
\end{equation}
where $\sigma_{ab}$ stands for the $1\sigma$ uncertainty on $\Gamma_{ab}$ arising from the cosmic variance. 
Here, we consider a fiducial model with massless gravitons, indicating $v_{g}=1$ during derivations of the Fisher matrix. 
By evaluating $F^{-1}$, we obtain the cosmic-variance limit on the speed of gravitational waves, i.e., $\sigma_{v_{g}}^{\mathrm{CV}}=0.07$. 
By using Eq.~(\ref{eq:vgmg}), we recast it into the cosmic-variance limit on the graviton mass bounds, namely,
\begin{equation}
    \sigma_{m_{g}}^{\mathrm{CV}} = 4.8 \times10^{-24} \mathrm{eV} \times \frac{f}{(10~\mathrm{year})^{-1}}\ ,
\end{equation}
where $f$ typically denotes the lowest frequency band of \ac{PTA} observations, which can roughly correspond to the observational time span. 
This inevitable uncertainty can not be gotten rid of by any \ac{PTA}, indicating that it is intrinsic. 
In other words, it stands for the maximal capability of \ac{PTA} in measuring the graviton mass.

\section{Conclusions and discussion}\label{sec:conclusion}

In this work, we investigated the graviton mass bounds through analysis of the \ac{NG15} 15-year dataset and the \ac{CPTA} first data release. 
By analyzing the data points of \acp{ORF} from these two observatories, we inferred the allowed parameter intervals for the speed of gravitational waves, particularly, its posterior probability distributions. 
By further recasting them into the posterior probability distributions of graviton mass, we obtained the state-of-the-art upper limit on the graviton mass, namely, $m_g\lesssim8.6\times10^{-24}\mathrm{eV}$ at 90\% confidence level.

We found that these new bounds on the graviton mass are comparable to the most recent one reported by the \ac{LVK} Collaborations, i.e., $ m_g < 1.27 \times 10^{-23} \mathrm{eV}$ at 90\% confidence level \cite{LIGOScientific:2021sio}. 
However, our study is solely based on the form of \acp{ORF}. 
Therefore, our work has two obvious advantages, compared with those works relevant to \ac{LVK}. 
The first one is independence on gravitational-wave sources. 
The second one is no requirement of prior knowledge of the waveforms. 
In addition, the findings observed by \ac{PTA} could effectively supplement those recorded by \ac{LVK}, since they could compose a multi-band study of the graviton mass.

Furthermore, we investigated the cosmic-variance limit on the graviton mass bounds from any \ac{PTA} observations of \ac{SGWB}, following an approach analogue to the study of \ac{CMB}. 
We found this inevitable cosmic-variance limit to be $\sigma_{m_{g}}^{\mathrm{CV}}=4.8\times10^{-24}\mathrm{eV}\times f/(10\mathrm{year})^{-1}$, with $f$ being the typical frequency band of \ac{PTA} observations of \ac{SGWB}. 
As an intrinsic uncertainty, it restricts the maximal capability of \ac{PTA} in measuring the graviton mass. 
Moreover, we also obtained the cosmic-variance limit on the speed of gravitational waves, i.e., $\sigma_{v_{g}}^{\mathrm{CV}}=0.07$.

One may wonder why the results on the \ac{NG15} data does not look consistent with those of a subsequent work \cite{Bi:2023ewq}, which adopts the same data to constrain the \ac{GW} speed. 
There might be several potential reasons to account for this discrepancy. 
A notable reason is the inclusion of cosmic variance in our analysis, a factor that was not accounted for in that work. 
The other reason involves an inherent difference between our model and that of Ref.~\cite{Bi:2023ewq}, leading to, e.g., different priors for the two works.

There might be implications of our study for scenarios of ultralight tensor dark matter \cite{Marzola:2017lbt,Hassan:2011zd}, which could account for the mystery of dark matter in the universe. 
In particular, the ultralight dark matter in the mass range $m_{\mathrm{uldm}}\sim10^{-22}\mathrm{eV}$ was used for conquering several shortcomings of the traditional cold dark matter \cite{Hu:2000ke,Hui:2016ltb}. 
To some extent, the ultralight dark matter behaves like the massive gravitons, indicating possible imprints of it on \ac{PTA} observations \cite{Broadhurst:2023tus,Konoplya:2023fmh,Yang:2023aak}. 
Since our results have been obtained through analysis of the recent \ac{PTA} datasets, we can infer that the preferred mass range of ultralight tensor dark matter is also compatible with the upper bounds on graviton mass shown by our current work. 
A related work resulting from detailed analysis of the \ac{NG15} 12.5-year dataset can be found in Ref.~\cite{Wu:2023dnp}.

\begin{acknowledgements}
We acknowledge Dr.~R.~C.~Bernardo for helpful discussion. 
S.W. is partially supported by the National Natural Science Foundation of China (Grant No. 12175243), the National Key R\&D Program of China No. 2023YFC2206403, the Science Research Grants from the China Manned Space Project with No. CMS-CSST-2021-B01, and the Key Research Program of the Chinese Academy of Sciences (Grant No. XDPB15). Z.C.Z. is supported by the National Key Research and Development Program of China Grant No. 2021YFC2203001. This work is supported by the High-performance Computing Platform of China Agricultural University. 
\end{acknowledgements}

%%%%%%%%%%%%%%%%%%%%%%%%%%%%%%%%%%%%%%%%%%%%%%%

\bibliography{pgw}
\bibliographystyle{JHEP}

\end{document}